\documentstyle[12pt]{article}
\newlength{\extraspace}
\newlength{\extraspaces}
\newcommand{\be}{\begin{equation}
\addtolength{\abovedisplayskip}{\extraspaces}
\addtolength{\belowdisplayskip}{\extraspaces}
\addtolength{\abovedisplayshortskip}{\extraspace}
\addtolength{\belowdisplayshortskip}{\extraspace}}
\newcommand{\ee}{\end{equation}}
\newcommand{\ba}{\begin{eqnarray}
\addtolength{\abovedisplayskip}{\extraspaces}
\addtolength{\belowdisplayskip}{\extraspaces}
\addtolength{\abovedisplayshortskip}{\extraspace}
\addtolength{\belowdisplayshortskip}{\extraspace}}
\newcommand{\ea}{\end{eqnarray}}
\newcommand{\nonu}{\nonumber \\[.5mm]}
\newcommand{\A}{&\!\!\!}
\begin{document}
\thispagestyle{empty}
\setlength{\baselineskip}{6mm}
\begin{flushright}
SIT-LP-11/04 \\
April, 2011
\end{flushright}
\vspace{7mm}
\begin{center}
{\large \bf Note on SUSY BF theory in $(1+2)$ dimensions \\[2mm]
from SUSY algebra for vector-spinor field
} \\[20mm]
{\sc Kazunari Shima}
\footnote{
\tt e-mail: shima@sit.ac.jp} \ 
and \ 
{\sc Motomu Tsuda}
\footnote{
\tt e-mail: tsuda@sit.ac.jp} 
\\[5mm]
{\it Laboratory of Physics, 
Saitama Institute of Technology \\
Fukaya, Saitama 369-0293, Japan} \\[20mm]
\begin{abstract}
We show in $(1+2)$ dimensions that supersymmetric (SUSY) $BF$ action 
for a (minimal and off-shell) spin-$\left( 1, {3 \over 2} \right)$ supermultiplet 
is a unique SUSY invariant one required from the closure property of commutator algebra for SUSY transformations. 
\\[5mm]
\end{abstract}
\end{center}

\newpage

\noindent
As a toy model to study the genuine topological structures of the field theory model, 
$BF$ theory and its supersymmetric (SUSY) version \cite{BG} are considered. 
In this short note we show another meaning of SUSY $BF$ theory 
containing a vector-spinor (high-spin) field in $(1+2)$ dimensions 
from the SUSY algebra for a spin-$\left( 1, {3 \over 2} \right)$ supermultiplet. 

Let us consider a set of the {\it minimal} and {\it off-shell} component fields $(v_a, \lambda_a, B_a)$, 
where we denote $v_a$ for a ($U(1)$) vector field, $\lambda_a$ for a (Majorana) vector-spinor field 
and $B_a$ for a auxiliary vector field, respectively. 
We discuss the closure property of commutator algebra 
for the most general form of SUSY \cite{GL}-\cite{WZ2} transformations in $d = 3$, 
\ba
\delta_\zeta v_a \A = \A a_1 \bar\zeta \lambda_a + i a_2 \epsilon_{abc} \bar\zeta \gamma^b \lambda^c, 
\nonu
\delta_\zeta \lambda_a \A = \A i b_1 F_{ab} \gamma^b \zeta + b_2 \epsilon_{abc} F^{bc} \zeta 
+ b_3 B_a \zeta + i b_4 \epsilon_{abc} B^b \gamma^c \zeta, 
\nonu
\delta_\zeta B_a \A = \A i c_1 \bar\zeta \!\!\not\!\partial \lambda_a 
+ i c_2 \bar\zeta \gamma_a \partial \cdot \lambda 
+ i c_3 \bar\zeta \partial_a \!\!\not\!\lambda 
+ c_4 \epsilon_{abc} \bar\zeta \partial^b \lambda^c, 
\label{genSUSY}
\ea
where $F_{ab} = \partial_a v_b - \partial_b v_a$ is the $U(1)$ gauge field strength, 
$\zeta$ is a constant (Majorana) spinor parameter for the SUSY transformations 
and $\{ a_i, b_i, c_i \}$ are arbitrary (real) parameters. 

The off-shell closure property of the commutator algebra for the SUSY transformations (\ref{genSUSY}) 
up to terms for gauge transformations requires 
\ba
\A \A 
a_2 = b_2 = b_4 = c_2 = c_4 = 0, 
\label{para1}
\\
\A \A 
a_1 b_1 + b_3 c_1 = 0. 
\label{para2}
\ea
The condition (\ref{para1}) restricts the form of Eq.(\ref{genSUSY}) to 
\ba
\delta_\zeta v_a \A = \A a_1 \bar\zeta \lambda_a, 
\nonu
\delta_\zeta \lambda_a \A = \A i b_1 F_{ab} \gamma^b \zeta + b_3 B_a \zeta, 
\nonu
\delta_\zeta B_a \A = \A i c_1 \bar\zeta \!\!\not\!\partial \lambda_a 
+ i c_3 \bar\zeta \partial_a \!\!\not\!\lambda, 
\label{SUSY}
\ea
which produce the closed off-shell commutator algebra 
\be
[ \delta_{\zeta_1}, \delta_{\zeta_2} ] 
= \delta_P(\Xi^a) + \delta_g(\theta) + \delta_g(\varepsilon) + \delta_g(\eta), 
\ee
where $\{ \delta_P(\Xi^a), \delta_g(\theta), \delta_g(\varepsilon), \delta_g(\eta) \}$ are 
\ba
\A \A 
\delta_P(\Xi^a) \ ... \ {\rm translations\ with\ a\ parameter\ } 
\Xi^a = 2i a_1 b_1 \bar\zeta_1 \gamma^a \zeta_2, 
\nonu
\A \A 
\delta_g(\theta) \ ... \ {\rm the\ } U(1)\ {\rm gauge\ transformation\ with\ a\ parameter\ } 
\theta = - 2i a_1 b_1 \bar\zeta_1 \gamma^a \zeta_2 v_a, 
\nonu
\A \A 
\delta_g(\varepsilon) \ ... \ {\rm a\ spinor\ gauge\ transformation\ with\ a\ parameter\ } 
\nonu
\A \A 
\hspace{1.5cm} \varepsilon = - i (a_1 b_1 + b_3 c_3) \bar\zeta_1 \gamma^b \zeta_2 \lambda_b 
+ (a_1 b_1 - b_3 c_3) \epsilon_{bcd} \bar\zeta_1 \gamma^b \zeta_2 \gamma^c \lambda^d, 
\nonu
\A \A 
\delta_g(\eta) \ ... \ {\rm a\ gauge\ transformation\ with\ a\ parameter\ } 
\nonu
\A \A 
\hspace{1.5cm} \eta = - 2i b_3 c_3 \bar\zeta_1 \gamma^a \zeta_2 B_a 
+ i b_1 (c_1 + 2 c_3) \epsilon_{bcd} \bar\zeta_1 \gamma^b \zeta_2 F^{cd}, 
\label{generator} 
\ea
respectively. 

Now we consider the invariant action. For (\ref{para2}) we put 
\be
{b_1 \over c_1} = {{-b_3} \over a_1} = \alpha, 
\ee
and starting from the most general Lorentz invariant action for the minimal off-shell supermultiplet, 
we find the following SUSY and gauge invariant action which is unique under the SUSY transformations (\ref{SUSY}), 
\be
L = {1 \over 2} \alpha \epsilon^{abc} B_a F_{bc} 
+ {1 \over 2} \epsilon^{abc} \bar\lambda_a \partial_b \lambda_c. 
\label{BFaction}
\ee
This is SUSY $BF$ action for $\alpha = 1$ in $d = 3$ space-time, 
which was first obtained as the SUSY extension of ordinary $BF$ theory \cite{BG}. 
We just mention that when $c_3 = - c_1$ with the condition (\ref{para2}) 
the SUSY transformation $\delta_\zeta \lambda_a$ in Eq.(\ref{SUSY}) is written 
in terms of the spinor gauge invariant $R_{ab} = \partial_a \lambda_b - \partial_b \lambda_a$, 
and that the $\epsilon_{bcd} \bar\zeta_1 \gamma^b \zeta_2 \gamma^c \lambda^d$ terms vanish 
in the parameter $\varepsilon$ of Eq.(\ref{generator}). 

Therefore, we conclude that in $(1+2)$ dimensions the SUSY $BF$ action (\ref{BFaction}) 
for the (minimal and off-shell) spin-$\left( 1, {3 \over 2} \right)$ supermultiplet 
is the unique SUSY invariant one required from the closure property of the SUSY transformations. 
These results seem characteristic in contrast with the situation for lower spin supermultiplets, 
for fermionic physical (though free) and bosonic unphysical degrees of freedom would form supermultiplet, 
which may give new insights to the role of high spin fields \cite{dWF} in the (local) SUSY theory model. 
Similar arguments in $d = 4$ for higher spin fields are future problems.

\newpage

%
\newcommand{\NP}[1]{{\it Nucl.\ Phys.\ }{\bf #1}}
\newcommand{\PL}[1]{{\it Phys.\ Lett.\ }{\bf #1}}
\newcommand{\CMP}[1]{{\it Commun.\ Math.\ Phys.\ }{\bf #1}}
\newcommand{\MPL}[1]{{\it Mod.\ Phys.\ Lett.\ }{\bf #1}}
\newcommand{\IJMP}[1]{{\it Int.\ J. Mod.\ Phys.\ }{\bf #1}}
\newcommand{\PR}[1]{{\it Phys.\ Rev.\ }{\bf #1}}
\newcommand{\PRL}[1]{{\it Phys.\ Rev.\ Lett.\ }{\bf #1}}
\newcommand{\PTP}[1]{{\it Prog.\ Theor.\ Phys.\ }{\bf #1}}
\newcommand{\PTPS}[1]{{\it Prog.\ Theor.\ Phys.\ Suppl.\ }{\bf #1}}
\newcommand{\AP}[1]{{\it Ann.\ Phys.\ }{\bf #1}}

\end{document}